\documentclass[aps,prb,10pt,twocolumn,amsmath,amssymb,superscriptaddress]{revtex4-2}

\usepackage{graphicx}
\usepackage{amsmath}
\usepackage{dcolumn}
\usepackage{bm}
\usepackage{xcolor}
\usepackage{graphicx}
\usepackage{physics}
\usepackage{mathptmx}
\usepackage{lipsum}
\usepackage{subfigure}
\usepackage{fixltx2e}
\usepackage[version=4]{mhchem} 
\usepackage[cal = cm]{mathalpha} 
\begin{document}

\title{Two-dimensional altermagnets from high throughput computational screening: symmetry requirements, chiral magnons and spin-orbit effects}

\author{Joachim Sødequist}
 \affiliation{Computational Atomic-Scale Materials Design (CAMD), Department of Physics, Technical University of Denmark, 2800 Kgs. Lyngby, Denmark}
\author{Thomas Olsen}
\email{tolsen@fysik.dtu.dk}
\affiliation{Computational Atomic-Scale Materials Design (CAMD), Department of Physics, Technical University of Denmark, 2800 Kgs. Lyngby, Denmark}

\date{\today}

\begin{abstract}
We present a high throughput computational search for altermagnetism in two-dimensional (2D) materials based on the Computational 2D Materials Database (C2DB). We start by showing that the symmetry requirements for altermagnetism in 2D are somewhat more strict compared to bulk materials and applying these yields a total of 7 altermagnets in the C2DB. The collinear ground state in these monolayers are verified by spin spiral calculations using the generalized Bloch theorem. We focus on four $d$-wave altermagnetic materials belonging to the $P2_1'/c'$ magnetic space group - RuF$_4$, VF$_4$, AgF$_2$ and OsF$_4$. The first three of these are known experimentally as van der Waals bonded bulk materials and are likely to be exfoliable from their bulk parent compounds. We perform a detailed analysis of the electronic structure and non-relativistic spin splitting in $k$-space exemplified by RuF$_4$. The magnon spectrum of RuF$_4$ is calculated from the magnetic force theorem and it is shown that the symmetries that enforce degenerate magnon bands in anti-ferromagnets are absent in altermagnets and give rise to the obtained non-degenerate magnon spectrum. We then include spin-orbit effects and show that these will dominate the splitting ofmagnons in RuF$_4$. Finally, we provide an example of $i$-wave altermagnetism in the 2H phase of FeBr$_3$. 
\end{abstract}

\maketitle



There has recently been considerable interest in the concept of altermagnetism \cite{hayami2019momentum,yuan2020giant,vsmejkal2022anomalous,cui2023efficient}, which denotes magnetic order with perfect spin-compensated magnetic order and non-relativistic spin-split bands. While the real space magnetic order resembles collinear antiferromagnetism the spin-splitting in $k$-space constitutes a natural generalisation of ferromagnetism. Moreover, since the spin-splitting is not governed by relativistic effects, the magnitude of the splitting may become much larger than typical degeneracy breaking originating from spin-orbit interactions in low-$Z$ elements. Compared to antiferromagnets, the spin-split bands in altermagnets make it possible to control the individual spin channels - without the complications of the macroscopic magnetisation and associated stray fields intrinsically present in ferromagnets - and renders them
promising for applications within spintronics, superconductivity and chiral magnonics \cite{cui2023efficient,vsmejkal2022emerging,vsmejkal2022chiral}. 
Several specific materials have already been reclassified from antiferromagnetic to altermagnetic. In particular, the well-known antiferromagnetic rutile structures, MnF$_2$\cite{yuan2020giant} and RuO$_2$\cite{vsmejkal2022chiral} as well as several new materials found by high throughput screening \cite{gao2023ai}.

In parallel with this, the discovery of ferromagnetic order in a monolayer of CrI$_3$ in 2017 \cite{huang2017layer} has spurred a booming interest in 2D magnetism. 
Since then, several monolayers have been added to the portfolio of 2D magnetic materials. For example, the itinerant ferromagnetic FeGeTe$_3$\cite{fei2018two}, antiferromagnetic FePS$_3$ \cite{wang2016raman,lee2016ising} and the spiral ordered NiI$_2$ \cite{Song2022} to name a few. In addition to this, there has been theoretical predictions of altermagnetism in specific 2D materials such as V$_2$Se$_2$O \cite{ma2021multifunctional}, V$_2$Te$_2$O \cite{cui2023giant}, Cr$_2$Te(Se)$_2$O \cite{cui2023efficient}, Cr$_2$O$_2$ \cite{chen2023giant,guo2023quantum}, and Cr$_2$SO\cite{guo2023piezoelectric}.
In 2D the Mermin-Wagner theorem forbids spontaneous symmetry breaking \cite{mermin1966absence,halperin2019hohenberg} and magnetic order thus has to be driven by magnetic anisotropy originating from spin-orbit coupling (SOC). Although, the spin-splitting in altermagnets have a non-relativistic origin, the magnetic order itself depends on SOC and the critical temperature of 2D magnets is in general expected to scale with the strength of SOC. It is therefore crucial to include spin-orbit effects in 2D magnets and it is presently not clear whether or not non-relativistic (proper altermagnetic) spin splitting will be important in materials with strong SOC.

\begin{figure*}[tb]
    \centering
    \includegraphics[width=\linewidth]{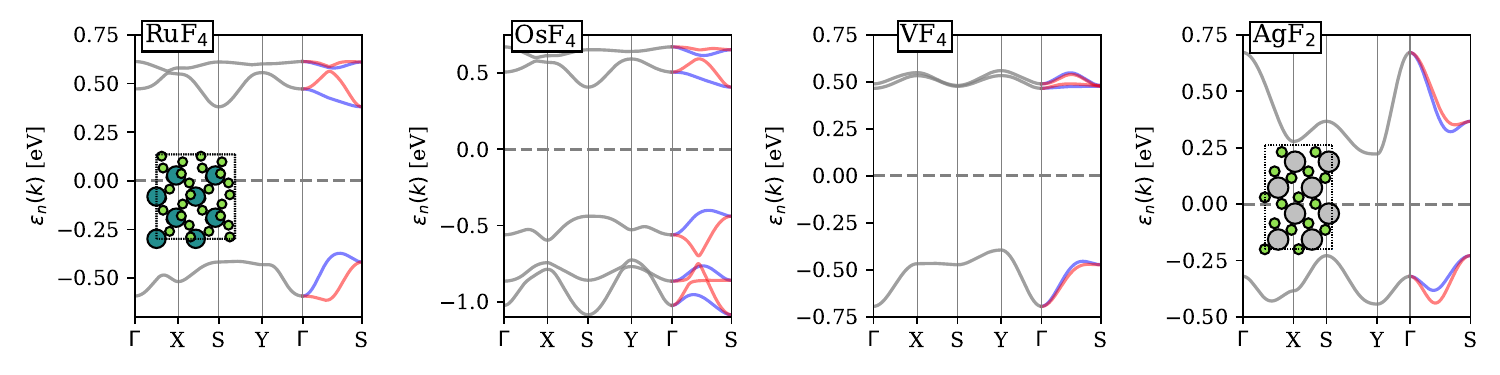}
    \caption{The band structures of altermagnetic RuF$_4$, OsF$_4$, VF$_4$ and AgF$_2$ calculated without SOC. The blue and red lines correspond to the two spin channels. The insets show the atomic structures in 2x2 supercells (OsF$_4$ and VF$_4$ are isostructural to RuF$_4$).}
    \label{fig:overview}
\end{figure*}
\begin{table}[b]
\begin{tabular}{l|c|c|c|c}
                 & SG  & MSG w/o SOC         & MSG w SOC          & Easy axis  \\ \hline
AgF$_2$          & $P2_1/c$ & $P2_1'/c'$      & $P2_1'/c'$      & $\hat{x}$  \\
RuF$_4$          & $P2_1/c$ & $P2_1'/c'$      & $P2_1/c$        & $\hat{y}$  \\
VF$_4$           & $P2_1/c$ & $P2_1'/c'$      & $P2_1/c$        & $\hat{y}$  \\
OsF$_4$          & $P2_1/c$ & $P2_1'/c'$      & $P2_1/c$        & $\hat{y}$  \\
FeBr$_3$         &  $P\Bar{6}2m$ & $P\Bar{6}2'm'$  & $P\Bar{6}2m$    & $\hat{z}$  \\
V$_2$ClBrI$_2$O$_2$  & $Cm$ & $Cm'$          & $P1$            & $\hat{x}$  \\
OsNNaSCl$_5$     &  $P2_1$ & $P2_1'$         &  $P2_1$          & $\hat{x}$ 
\end{tabular}
\caption{List of altermagnetic monolayers in the C2DB. We state the space groups (SG), magnetic space groups (MSG) with and without SOC, and the easy axis.}
\label{tab:altermagnets}
\end{table}

If SOC is excluded one may freely perform a global rotation of the spin without changing the energy. In particular one may consider a rotation $U$, which flips the sign of the spin. In addition, for collinear structures, the mean-field Hamiltonian commutes with complex conjugation ($K$) and in $k$-space eigenstates at $\pm \mathbf{k}$ in will be degenerate. Thus we may consider the time-reversal operator $\theta=KU$, which act as inversion in $k$-space (while flipping the spin), but has the same effect as $U$ in real space. In addition, since the direction of spin is arbitrary it may be regarded as invariant under any unitary symmetry operation (in contrast to anti-unitary transformations involving $\theta$). Alternatively, one may regard the spin as an axial vector along some chosen direction and then cancel the effect of rotations by a suitable global spin rotation. 
The possible  magnetic space groups (MSGs) that can host compensated magnetic order are then limited to those of type III and type IV  - often called the black-and-white MSGs. In 3D there are two fundamental symmetries, which enforce spin-degeneracy of the band structure \cite{yuan2020giant}. The first is inversion symmetry ($I$) combined with time reversal, $\{I|\mathbf{0}\}\theta$. The second is the combination of fractional translation $\boldsymbol{\tau}$ and a spinor rotation $\{E | \boldsymbol{\tau}\}U$ \cite{yuan2020giant}. 
Since all type IV MSGs have $\{E | \boldsymbol{\tau}\}U$ symmetry, spin-splitting can only occur in type IV MSGs as a consequence of SOC. Equivalently, these conditions can be identified as those of the type II spin-group of non-relativistic collinear magnetism \cite{vsmejkal2022beyond}. 
In 2D materials, there are additional symmetries that leave the 2D Brillouin zone invariant. If we assume the monolayer is spanned by the $xy$-plane, a two-fold rotation axis along $z$ combined with time-reversal $\{C_{2z}| \mathbf{0}\}\theta$, will effectively act as inversion in $k$-space. 
In addition, out-of-plane mirror symmetry combined with a global spinor rotation $\{m_z|\mathbf{0}\}U$ may also prohibit spin-splitting since the mirror must interchange two atoms of opposite spin. 
The 2D $k$-space is left invariant, thus enforcing spin degeneracy in the entire 2D Brillouin zone. This symmetry may be extended to glide planes $\{m_z|\boldsymbol{\tau}\}U$ by the same argument. 
In line with previous work on the topic we consider altermagnets as being materials with compensated collinear magnetic order and absence of these symmetries in the MSG when SOC is excluded \cite{vsmejkal2022emerging}.

\begin{figure*}[tb]
    \centering
    \includegraphics[width=\linewidth]{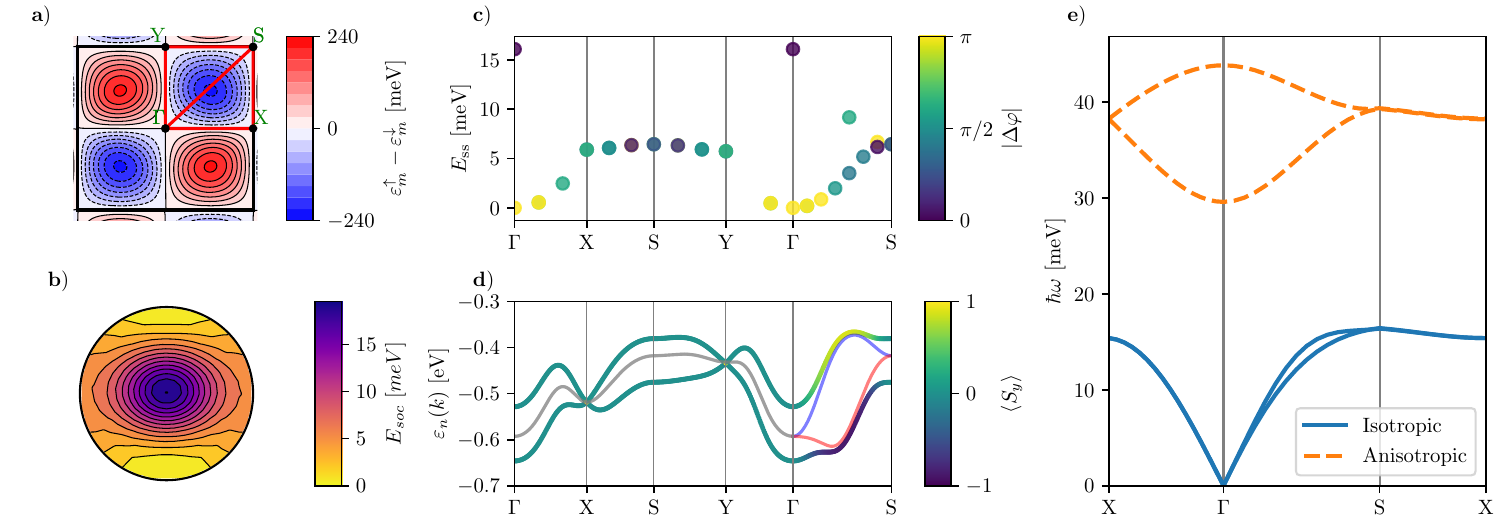}
    \caption{Electronic and magnetic properties of altermagnetic RuF$_4$. a) The spin splitting of the two highest occupied bands calculated without SOC. b) The spin-orbit energy as a function of spin orientation on the upper hemisphere shown as a stereographic projection. c) Spin spiral dispersion of RuF$_4$ coloured by the angle between the two magnetic moments in the unit cell. d) The electronic band structure with and without SOC. The bands including SOC are coloured according to $\langle S_y\rangle$ in units of $\hbar/2$ e) Magnon dispersion excluding (isotropic) and including (anisotropic) SOC.}
    \label{fig:ruf4}
\end{figure*}
The starting point of the present work is the C2DB, which contains 1681 ferromagnetic monolayers relaxed using the Perdew–Burke-Ernzerhof (PBE) exchange-correlation functional \cite{Gjerding2021}. The true magnetic ground state of these materials has only been calculated rigorously for monolayers with one magnetic atom in the chemical unit cell \cite{sodequist2023magnetic} and all such compounds cannot have a MSG of type III, which disqualifies them as altermagnetic. They may, however, still have a collinear ground state with spin-splitting driven by SOC and as shown in Ref. \cite{sodequist2023magnetic} the C2DB presently contains 17 such compounds.
For magnetic materials with two magnetic moments in the unit cell, the energy difference between the ferromagnetic state and the spin-compensated state has previously been calculated \cite{torelli2020first} and we can restrict the set of candidate materials to those where the spin-compensated state has lower energy. This yields an initial set of 170 materials, which are predicted to have a non-ferromagnetic ground state and whether or not the collinear compensated structure in the minimal unit cell is the true ground state has to be verified from spin spiral calculations as described below.  
For the spin-compensated state of a given material we used the python package spglib \cite{spglibv1,spglibv2} to extract the MSG without SOC. The symmetries of the determined MSG are then obtained and the material is discarded as an altermagnetic candidate if either $\{I|\mathbf{0}\}\theta$, $\{C_{2z}|\mathbf{0}\}\theta$, $\{m_z|\mathbf{0}\}U$ or $\{m_z|\boldsymbol{\tau}\}U$ is present.
We note that the symmetry analysis requires some care since the determined symmetries of a material relaxed with density functional theory will depend on a certain tolerance as well as the accuracy with which the atomic positions were relaxed. 
In order to ensure that symmetries and the associated spin splitting in a candidate material are not due to inaccurate relaxation, we performed additional symmetrised relaxation. For a given material we thus carried out an initial unconstrained relaxation and then determined all unique space groups within any symmetry tolerance. For that material a set of symmetrised structures that conform to the different possible space groups was then generated and all structures were relaxed under constraint of the associated symmetries. Finally, we picked the structure with lowest energy, which is then guaranteed to fulfil the symmetries of the determined MSG exactly.
For the predicted altermagnets we verified that the bands exhibit alternating spin-splitting in reciprocal space and used spin spiral calculations to verify that the collinear altermagnetic state constitutes the true magnetic ground state \cite{sodequist2023type,kurz2004ab,knopfle2000spin}. 
All calculations presented here were carried out using the electronic structure code GPAW \cite{Enkovaara2010, mortensen2023gpaw}, which is based on the projector-augmented wave method. We used a plane wave basis with a cutoff of 800 eV and the PBE functional for relaxation and band structures.

The workflow described above yields a total of 7 altermagnetic monolayers in the C2DB, 
which are listed in table \ref{tab:altermagnets}. Four of these materials belong to the space group $P2_1/c$ and are all structurally similar fluorides with rectangular unit cells: RuF$_4$, VF$_4$, AgF$_2$ and OsF$_4$. The first three of these are experimentally known in stable van der Waals bonded bulk crystal and thus have an associated ICSD \cite{Allmann2007} or COD \cite{Graulis2011} identifier and are likely to be exfoliable from the bulk parent compounds.
The band structures without SOC are shown in figure \ref{fig:overview} and all compounds are semiconducting with band gaps of roughly 1 eV.
The spin-splitting is found to depend strongly on the magnetic element; in RuF$_4$ and OsF$_4$ the maximal spin-splitting is 240 meV while in the isostructural VF$_4$ it is 50 meV.
The spin-compensated structures have the MSGs $P2_1'/c'$ (excluding SOC). This contains two unitary symmetries $G_{U} = \left(\{E | \mathbf{0}\}, \{I | \mathbf{0}\}\right)$ and two anti-unitary symmetries $G_{AU} = \left(\{C_{2x} | \boldsymbol{\tau}_x\}\theta, \{m_x | \boldsymbol{\tau}_y\}\theta\right)$, where $\boldsymbol{\tau}_{x/y}$ is half a lattice vector. The bands are generally non-degenerate, but spin-degeneracy is enforced on all high symmetry lines (except $\Gamma$S) due to the anti-unitary screw axis and glide plane. 

In figure \ref{fig:ruf4}a we show the spin-splitting of the two highest valence bands of RuF$_4$.
The spin-splitting is seen to have $d$-wave symmetry, with the sign of the splitting alternating twice for any path encircling the $\Gamma$-point. We found a similar symmetry of the spin-splitting in all of the 2D altermagnets except FeBr$_3$, which will be discussed below.
The magnetic ground state of the altermagnets has been verified using spin spiral calculations. We found all materials to have ground states with $\mathbf{Q}=\Gamma$ with compensated collinear magnetic moments in the unit cell. We exemplify one of these calculations for RuF$_4$ in figure \ref{fig:ruf4}c. The DFT total energy is calculated using the Local Density Approximation (LDA) functional as a function of the magnetic order vector $\mathbf{q}$ along the high symmetry path of the crystal. The ground state is seen to be located at $\mathbf{Q}=\Gamma$ with anti-parallel intra-cell magnetic order, which corresponds to a standard Néel type order. 


When SOC is included the magnetisation acquires a preferred direction. In figure \ref{fig:ruf4}b we show the SOC energy as a function of spin-direction and the easy-axis is determined to be along the $y$-direction. The MSG including SOC (regarding magnetisation as an axial vector) is then found to be $P2_1/c$, which has no anti-unitary symmetries and consequently, spin-splitting is expected in the entire Brillouin zone. In figure \ref{fig:ruf4}d, we show the band structure including SOC, which shows that the two highest lying valence bands become split along most of the high symmetry path, but the projection of the spin along $y$ vanishes along the entire path (except for the $\Gamma$S segment) as a consequence of the unitary glide plane and screw axis. 
The degeneracies found in $X$ and $Y$ are a consequence of the non-symmorphic symmetries, which enforces at least one band crossing through the entire Brillouin zone \cite{young2015dirac,zhao2016nonsymmorphic}. We note that when SOC is included the MSG depends on the easy-axis and AgF$_2$ thus acquires a different MSG (type III) than the remaining $P2_1/c$ compounds (see table \ref{tab:altermagnets}).

\begin{figure}[tb]
    \centering
    \includegraphics[width=1\linewidth]{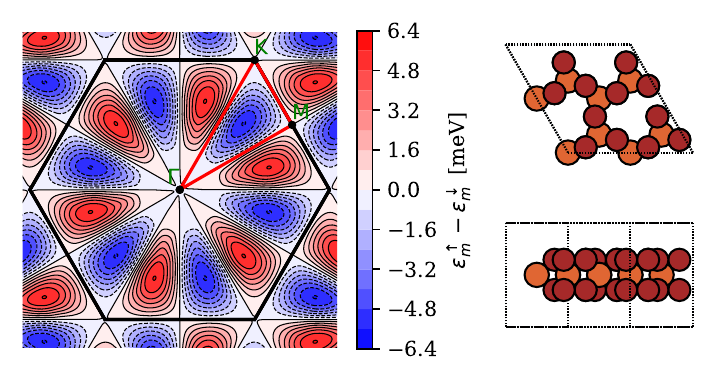}
    \caption{Altermagnetic $i$-wave spin-splitting of the two highest occupied bands in 2H-FeBr$_3$.}
    \label{fig:iwave}
\end{figure}

The spin compensation and inversion symmetry in $k$-space 
imply that the spin splitting in altermagnets may have either have $d$-, $g$- or $i$-wave symmetry. For the 7 altermagnetic 2D compounds found here, 6 have $d$-wave symmetry while one case - FeBr$_3$ - has $i$-wave symmetry. The spin splitting of the two highest valence bands of FeBr$_3$ is shown in figure \ref{fig:iwave}, and exhibits 12 polarisation reversals for any path encircling the $\Gamma$-point. 
The path $\Gamma$K is spin degenerate as a consequence of a mirror plane orthogonal to the atomic plane containing the $(11)$-direction (combined with a global spinor rotation), while the degeneracies at the $\Gamma$M and MK segments follow from other mirror planes combined with $\theta$.

In order to analyse the magnon dispersion of altermagnets we consider the isotropic Heisenberg model
\begin{align}\label{eq:H}
H=-\frac{1}{2}\sum_{abij}J_{abij}\mathbf{S}_{ai}\cdot\mathbf{S}_{bj},
\end{align}
where $\mathbf{S}_{ai}$ is the spin operator for magnetic atom $a$ in unit cell $i$ and $J_{abij}$ are the exchange couplings determining the magnetic interactions between sites. It is straightforward to apply semi-classical spin-wave theory and show that the magnon energies are given by the positive eigenvalues of the dynamical matrix
\begin{align}\label{eq:H_dyn}
H_{ab}(\mathbf{q})=
\begin{bmatrix}
0 & B_{ab}(\mathbf{q})\\
C_{ab}(\mathbf{q}) & 0
\end{bmatrix},
\end{align}
with 
\begin{align}
&B_{ab}(\mathbf{q})=-iS_aJ_{ab}(\mathbf{q})+i\delta_{ab}\sum_cS_cJ_{ac}(\mathbf{0}),\label{eq:B}\\
&C_{ab}(\mathbf{q})=-B_{ab}(\mathbf{q}),\label{eq:C}
\end{align}
and 
\begin{align}
J_{ab}(\mathbf{q})=\sum_iJ_{abij}e^{i(\mathbf{R}_j-\mathbf{R}_i)\cdot\mathbf{q}},
\end{align}
where $\mathbf{R}_i$ is the lattice vector corresponding to unit cell $i$ and $S_a$ is the ground state spin of atom $a$ (maximum or minimum eigenvalue of $S^z_a$). One clearly has that $J_{abij}=J_{baji}$ and therefore $J_{ab}(-\mathbf{q})=J_{ab}^*(\mathbf{q})=J_{ba}(\mathbf{q})$. The dynamical matrix can be made block diagonal by a unitary transformation and the magnon energies at a particular $\mathbf{q}$ can simply be determined from the magnitude of the eigenvalues of $B_{ab}(\mathbf{q})$. With two atoms in the unit cell the two magnon energies will then be degenerate unless $J_{11}(\mathbf{q})\neq J_{22}(\mathbf{q})$. 

In order to analyse the symmetry constraints on the exchange constants we note that these will transform as the reactive part of the static transverse magnetic susceptibility \cite{Durhuus2023}. In the absence of SOC a unitary symmetry $\mathcal{R}$ that leaves the ground state invariant implies \cite{Skovhus2022}
\begin{align}
\chi^{+-}(\mathbf{r},\mathbf{r}')=\chi^{+-}(\mathcal{R}\mathbf{r},\mathcal{R}\mathbf{r}'),
\end{align}
while an anti-unitary symmetry (complex conjugation combined with a unitary symmetry $\mathcal{R}$) yields \cite{Skovhus2022}
\begin{align}
\chi^{+-}(\mathbf{r},\mathbf{r}')=\chi^{+-}(\mathcal{R}\mathbf{r}',\mathcal{R}\mathbf{r}).
\end{align}
This implies in particular that anti-unitary symmetries that exchange sublattices yield
\begin{align}\label{eq:sym_J}
J_{11}(\mathbf{q})=J_{22}(-\mathcal{R}^T\mathbf{q}).
\end{align}
Specifically, for the common case of $I\theta$ symmetry one obtains $J_{11}(\mathbf{q})=J_{22}(\mathbf{q})$, which implies two degenerate magnon bands. The fact that one may have $J_{11}(\mathbf{q})\neq J_{22}(\mathbf{q})$ despite the fact that magnetic moments at the two sites are related by symmetry is a peculiar feature of altermagnets, which renders the magnon spectrum non-degenerate when SOC is excluded from the analysis. The isotropic model will always exhibit reciprocity such that any magnon energy at $\mathbf{q}$ will be degenerate
with one at $-\mathbf{q}$. These states will, however, exhibit opposite chirality \cite{vsmejkal2022chiral}, whereas the degenerate magnon bands in antiferromagnets will not have well-defined chirality (when SOC is excluded). The magnon spectrum of RuF$_4$ -- modelled by an isotropic Heisenberg Hamiltonian -- is shown in figure \ref{fig:ruf4}e and is seen to exhibit split magnons bands along the $\Gamma$S-direction. The magnons are in fact non-degenerate in the entire Brillouin zone except for the high symmetry lines $\Gamma$X, $\Gamma$Y, XS and XY, where either the glide plane or the two-fold screw axis enforces degeneracy through Eq. 
\eqref{eq:sym_J}. The exchange parameters of the model here was calculated from the magnetic force theorem \cite{Liechtenstein1987,Durhuus2023} using the LDA functional, a plane wave cutoff of 800 eV and a $30\times 30$ $k$-point grid.

Although the non-degenerate magnons of the isotropic Heisenberg model are interesting in their own right, the underlying non-relativistic assumption fails rather dramatically in the case of RuF$_4$. The simplest consequences of spin-orbit effects can be included through the addition of a single-ion anisotropy term in the model. The Hamiltonian \eqref{eq:H} is then augmented by the term
\begin{align}\label{eq:deltaH}
\Delta H^\mathrm{ani}=-\sum_{ia\alpha\beta}S_{ai}^\alpha A^{\alpha\beta}S_{ai}^{\beta}, 
\end{align}
where $A^{\alpha\beta}$ denotes the single-ion anisotropy tensor and $\alpha,\beta$ runs over the Cartesian directions. The tensor is symmetric and has three independent parameters corresponding to the eigendirections. For the case of RuF$_4$, the tensor is diagonal in the $x,y,z$ basis, with y being the easy-axis and $z$ (out-of-plane) the hard axis. The dynamical matrix \eqref{eq:H_dyn} -- \eqref{eq:C} then acquires the additional terms
\begin{align}
&\Delta B_{ab}(\mathbf{q})=-i\delta_{ab}\text{sign}(S_a)(2|S_a|-1)(A^{xx}-A^{yy}),\\
&\Delta C_{ab}(\mathbf{q})=-i\delta_{ab}\text{sign}(S_a)(2|S_a|-1)(A^{yy}-A^{zz}),
\end{align}
which in general will give rise to split magnon bands. The single-ion anisotropy tensor of RuF$_4$ was obtained by calculating the spin-orbit energy non-selfconsistently \cite{Olsen2016a, sodequist2023magnetic} for spins pointing along different directions $\mathbf{\hat n}$ (see figure \ref{fig:ruf4}b) and fitting the result to $\langle\mathbf{\hat n}|\Delta H| \mathbf{\hat n}\rangle$. We are free to set the zero point at the easy direction and then obtain $A^{yy}=0$, $A^{xx}=-5.7$ and $A^{zz}=-19.6$ meV. The exchange coupling is roughly given by half the magnon band width obtained without SOC and the anisotropy energy is thus seen to be more than twice the exchange energy. In figure \ref{fig:ruf4}e we show the magnon spectrum obtained from the anisotropic Heisenberg model. It is clear that the splitting is completely dominated by the anisotropy, while the altermagnetic splitting originating from bare exchange is of minor importance. The case of RuF$_4$ is a rather extreme example in this regard, since it exhibits rather weak exchange interactions as well as very strong spin-orbit effects. Typically, exchange interactions are much stronger than anisotropy parameters and the altermagnetic symmetries are then expected to play a more important role in the splitting of the magnon energies.

To summarise, we have screened the C2DB for altermagnetic 2D materials and found 7 altermagnets where one of them, 2H-FeBr$_3$, was found to exhibit $i$-wave altermagnetism, which to our knowledge, is the first of its kind.
Three of the structures in the $P2_1/c$ space group - RuF$_4$, VF$_4$ and AgF$_2$ - are derived from parent bulk compounds that have characterised experimentally and are thus promising candidates for realising 2D altermagnets. RuF$
_4$ exhibits the largest altermagnetic splitting, but we showed that SOC gives rise to spin-splitting of similar magnitude. This is due to the presence of Ru, which induces strong spin-orbit effects in this material and the situation may be different for materials containing lighter elements. However, strong SOC is a crucial prerequisite for magnetic order in 2D and it is not obvious that non-relativistic spin-splitting will ever dominate over spin-orbit induced spin-splitting in any 2D material with a sizeable Néel temperature. 
We finally calculated the magnetic excitation spectrum of RuF$_4$ using MFT and found that the altermagnetic magnon band splitting could be traced to the fact that the lack of symmetries does not enforce $J_{11}(\mathbf{q})=J_{22}(\mathbf{q})$ on the Heisenberg parameters. Although this is interesting from a fundamental point of view, it becomes irrelevant once spin-orbit effects are included, since the magnetic anisotropy will always yield magnon splitting in systems with tri-axial anisotropy. 

The authors acknowledge support from the Villum foundation Grant No. 00029378

\clearpage
\bibliography{bibliography.bib}

\begin{thebibliography}{38}%
\makeatletter
\providecommand \@ifxundefined [1]{%
 \@ifx{#1\undefined}
}%
\providecommand \@ifnum [1]{%
 \ifnum #1\expandafter \@firstoftwo
 \else \expandafter \@secondoftwo
 \fi
}%
\providecommand \@ifx [1]{%
 \ifx #1\expandafter \@firstoftwo
 \else \expandafter \@secondoftwo
 \fi
}%
\providecommand \natexlab [1]{#1}%
\providecommand \enquote  [1]{``#1''}%
\providecommand \bibnamefont  [1]{#1}%
\providecommand \bibfnamefont [1]{#1}%
\providecommand \citenamefont [1]{#1}%
\providecommand \href@noop [0]{\@secondoftwo}%
\providecommand \href [0]{\begingroup \@sanitize@url \@href}%
\providecommand \@href[1]{\@@startlink{#1}\@@href}%
\providecommand \@@href[1]{\endgroup#1\@@endlink}%
\providecommand \@sanitize@url [0]{\catcode `\\12\catcode `\$12\catcode `\&12\catcode `\#12\catcode `\^12\catcode `\_12\catcode `\%12\relax}%
\providecommand \@@startlink[1]{}%
\providecommand \@@endlink[0]{}%
\providecommand \url  [0]{\begingroup\@sanitize@url \@url }%
\providecommand \@url [1]{\endgroup\@href {#1}{\urlprefix }}%
\providecommand \urlprefix  [0]{URL }%
\providecommand \Eprint [0]{\href }%
\providecommand \doibase [0]{https://doi.org/}%
\providecommand \selectlanguage [0]{\@gobble}%
\providecommand \bibinfo  [0]{\@secondoftwo}%
\providecommand \bibfield  [0]{\@secondoftwo}%
\providecommand \translation [1]{[#1]}%
\providecommand \BibitemOpen [0]{}%
\providecommand \bibitemStop [0]{}%
\providecommand \bibitemNoStop [0]{.\EOS\space}%
\providecommand \EOS [0]{\spacefactor3000\relax}%
\providecommand \BibitemShut  [1]{\csname bibitem#1\endcsname}%
\let\auto@bib@innerbib\@empty
\bibitem [{\citenamefont {Hayami}\ \emph {et~al.}(2019)\citenamefont {Hayami}, \citenamefont {Yanagi},\ and\ \citenamefont {Kusunose}}]{hayami2019momentum}%
  \BibitemOpen
  \bibfield  {author} {\bibinfo {author} {\bibfnamefont {S.}~\bibnamefont {Hayami}}, \bibinfo {author} {\bibfnamefont {Y.}~\bibnamefont {Yanagi}},\ and\ \bibinfo {author} {\bibfnamefont {H.}~\bibnamefont {Kusunose}},\ }\bibfield  {title} {\bibinfo {title} {Momentum-dependent spin splitting by collinear antiferromagnetic ordering},\ }\href@noop {} {\bibfield  {journal} {\bibinfo  {journal} {journal of the physical society of japan}\ }\textbf {\bibinfo {volume} {88}},\ \bibinfo {pages} {123702} (\bibinfo {year} {2019})}\BibitemShut {NoStop}%
\bibitem [{\citenamefont {Yuan}\ \emph {et~al.}(2020)\citenamefont {Yuan}, \citenamefont {Wang}, \citenamefont {Luo}, \citenamefont {Rashba},\ and\ \citenamefont {Zunger}}]{yuan2020giant}%
  \BibitemOpen
  \bibfield  {author} {\bibinfo {author} {\bibfnamefont {L.-D.}\ \bibnamefont {Yuan}}, \bibinfo {author} {\bibfnamefont {Z.}~\bibnamefont {Wang}}, \bibinfo {author} {\bibfnamefont {J.-W.}\ \bibnamefont {Luo}}, \bibinfo {author} {\bibfnamefont {E.~I.}\ \bibnamefont {Rashba}},\ and\ \bibinfo {author} {\bibfnamefont {A.}~\bibnamefont {Zunger}},\ }\bibfield  {title} {\bibinfo {title} {Giant momentum-dependent spin splitting in centrosymmetric low-z antiferromagnets},\ }\href@noop {} {\bibfield  {journal} {\bibinfo  {journal} {Physical Review B}\ }\textbf {\bibinfo {volume} {102}},\ \bibinfo {pages} {014422} (\bibinfo {year} {2020})}\BibitemShut {NoStop}%
\bibitem [{\citenamefont {{\v{S}}mejkal}\ \emph {et~al.}(2022{\natexlab{a}})\citenamefont {{\v{S}}mejkal}, \citenamefont {MacDonald}, \citenamefont {Sinova}, \citenamefont {Nakatsuji},\ and\ \citenamefont {Jungwirth}}]{vsmejkal2022anomalous}%
  \BibitemOpen
  \bibfield  {author} {\bibinfo {author} {\bibfnamefont {L.}~\bibnamefont {{\v{S}}mejkal}}, \bibinfo {author} {\bibfnamefont {A.~H.}\ \bibnamefont {MacDonald}}, \bibinfo {author} {\bibfnamefont {J.}~\bibnamefont {Sinova}}, \bibinfo {author} {\bibfnamefont {S.}~\bibnamefont {Nakatsuji}},\ and\ \bibinfo {author} {\bibfnamefont {T.}~\bibnamefont {Jungwirth}},\ }\bibfield  {title} {\bibinfo {title} {Anomalous hall antiferromagnets},\ }\href@noop {} {\bibfield  {journal} {\bibinfo  {journal} {Nature Reviews Materials}\ }\textbf {\bibinfo {volume} {7}},\ \bibinfo {pages} {482} (\bibinfo {year} {2022}{\natexlab{a}})}\BibitemShut {NoStop}%
\bibitem [{\citenamefont {Cui}\ \emph {et~al.}(2023{\natexlab{a}})\citenamefont {Cui}, \citenamefont {Zeng}, \citenamefont {Cui}, \citenamefont {Yu},\ and\ \citenamefont {Yang}}]{cui2023efficient}%
  \BibitemOpen
  \bibfield  {author} {\bibinfo {author} {\bibfnamefont {Q.}~\bibnamefont {Cui}}, \bibinfo {author} {\bibfnamefont {B.}~\bibnamefont {Zeng}}, \bibinfo {author} {\bibfnamefont {P.}~\bibnamefont {Cui}}, \bibinfo {author} {\bibfnamefont {T.}~\bibnamefont {Yu}},\ and\ \bibinfo {author} {\bibfnamefont {H.}~\bibnamefont {Yang}},\ }\bibfield  {title} {\bibinfo {title} {Efficient spin seebeck and spin nernst effects of magnons in altermagnets},\ }\href@noop {} {\bibfield  {journal} {\bibinfo  {journal} {Physical Review B}\ }\textbf {\bibinfo {volume} {108}},\ \bibinfo {pages} {L180401} (\bibinfo {year} {2023}{\natexlab{a}})}\BibitemShut {NoStop}%
\bibitem [{\citenamefont {{\v{S}}mejkal}\ \emph {et~al.}(2022{\natexlab{b}})\citenamefont {{\v{S}}mejkal}, \citenamefont {Sinova},\ and\ \citenamefont {Jungwirth}}]{vsmejkal2022emerging}%
  \BibitemOpen
  \bibfield  {author} {\bibinfo {author} {\bibfnamefont {L.}~\bibnamefont {{\v{S}}mejkal}}, \bibinfo {author} {\bibfnamefont {J.}~\bibnamefont {Sinova}},\ and\ \bibinfo {author} {\bibfnamefont {T.}~\bibnamefont {Jungwirth}},\ }\bibfield  {title} {\bibinfo {title} {Emerging research landscape of altermagnetism},\ }\href@noop {} {\bibfield  {journal} {\bibinfo  {journal} {Physical Review X}\ }\textbf {\bibinfo {volume} {12}},\ \bibinfo {pages} {040501} (\bibinfo {year} {2022}{\natexlab{b}})}\BibitemShut {NoStop}%
\bibitem [{\citenamefont {{\v{S}}mejkal}\ \emph {et~al.}(2023)\citenamefont {{\v{S}}mejkal}, \citenamefont {Marmodoro}, \citenamefont {Ahn}, \citenamefont {Gonz{\'a}lez-Hern{\'a}ndez}, \citenamefont {Turek}, \citenamefont {Mankovsky}, \citenamefont {Ebert}, \citenamefont {D’Souza}, \citenamefont {{\v{S}}ipr}, \citenamefont {Sinova} \emph {et~al.}}]{vsmejkal2022chiral}%
  \BibitemOpen
  \bibfield  {author} {\bibinfo {author} {\bibfnamefont {L.}~\bibnamefont {{\v{S}}mejkal}}, \bibinfo {author} {\bibfnamefont {A.}~\bibnamefont {Marmodoro}}, \bibinfo {author} {\bibfnamefont {K.-H.}\ \bibnamefont {Ahn}}, \bibinfo {author} {\bibfnamefont {R.}~\bibnamefont {Gonz{\'a}lez-Hern{\'a}ndez}}, \bibinfo {author} {\bibfnamefont {I.}~\bibnamefont {Turek}}, \bibinfo {author} {\bibfnamefont {S.}~\bibnamefont {Mankovsky}}, \bibinfo {author} {\bibfnamefont {H.}~\bibnamefont {Ebert}}, \bibinfo {author} {\bibfnamefont {S.~W.}\ \bibnamefont {D’Souza}}, \bibinfo {author} {\bibfnamefont {O.}~\bibnamefont {{\v{S}}ipr}}, \bibinfo {author} {\bibfnamefont {J.}~\bibnamefont {Sinova}}, \emph {et~al.},\ }\bibfield  {title} {\bibinfo {title} {Chiral magnons in altermagnetic ruo 2},\ }\href@noop {} {\bibfield  {journal} {\bibinfo  {journal} {Physical Review Letters}\ }\textbf {\bibinfo {volume} {131}},\ \bibinfo {pages} {256703} (\bibinfo {year} {2023})}\BibitemShut {NoStop}%
\bibitem [{\citenamefont {Gao}\ \emph {et~al.}(2023)\citenamefont {Gao}, \citenamefont {Qu}, \citenamefont {Zeng}, \citenamefont {Wen}, \citenamefont {Sun}, \citenamefont {Guo},\ and\ \citenamefont {Lu}}]{gao2023ai}%
  \BibitemOpen
  \bibfield  {author} {\bibinfo {author} {\bibfnamefont {Z.-F.}\ \bibnamefont {Gao}}, \bibinfo {author} {\bibfnamefont {S.}~\bibnamefont {Qu}}, \bibinfo {author} {\bibfnamefont {B.}~\bibnamefont {Zeng}}, \bibinfo {author} {\bibfnamefont {J.-R.}\ \bibnamefont {Wen}}, \bibinfo {author} {\bibfnamefont {H.}~\bibnamefont {Sun}}, \bibinfo {author} {\bibfnamefont {P.}~\bibnamefont {Guo}},\ and\ \bibinfo {author} {\bibfnamefont {Z.-Y.}\ \bibnamefont {Lu}},\ }\bibfield  {title} {\bibinfo {title} {Ai-accelerated discovery of altermagnetic materials},\ }\href@noop {} {\bibfield  {journal} {\bibinfo  {journal} {arXiv preprint arXiv:2311.04418}\ } (\bibinfo {year} {2023})}\BibitemShut {NoStop}%
\bibitem [{\citenamefont {Huang}\ \emph {et~al.}(2017)\citenamefont {Huang}, \citenamefont {Clark}, \citenamefont {Navarro-Moratalla}, \citenamefont {Klein}, \citenamefont {Cheng}, \citenamefont {Seyler}, \citenamefont {Zhong}, \citenamefont {Schmidgall}, \citenamefont {McGuire}, \citenamefont {Cobden} \emph {et~al.}}]{huang2017layer}%
  \BibitemOpen
  \bibfield  {author} {\bibinfo {author} {\bibfnamefont {B.}~\bibnamefont {Huang}}, \bibinfo {author} {\bibfnamefont {G.}~\bibnamefont {Clark}}, \bibinfo {author} {\bibfnamefont {E.}~\bibnamefont {Navarro-Moratalla}}, \bibinfo {author} {\bibfnamefont {D.~R.}\ \bibnamefont {Klein}}, \bibinfo {author} {\bibfnamefont {R.}~\bibnamefont {Cheng}}, \bibinfo {author} {\bibfnamefont {K.~L.}\ \bibnamefont {Seyler}}, \bibinfo {author} {\bibfnamefont {D.}~\bibnamefont {Zhong}}, \bibinfo {author} {\bibfnamefont {E.}~\bibnamefont {Schmidgall}}, \bibinfo {author} {\bibfnamefont {M.~A.}\ \bibnamefont {McGuire}}, \bibinfo {author} {\bibfnamefont {D.~H.}\ \bibnamefont {Cobden}}, \emph {et~al.},\ }\bibfield  {title} {\bibinfo {title} {Layer-dependent ferromagnetism in a van der waals crystal down to the monolayer limit},\ }\href@noop {} {\bibfield  {journal} {\bibinfo  {journal} {Nature}\ }\textbf {\bibinfo {volume} {546}},\ \bibinfo {pages} {270} (\bibinfo {year} {2017})}\BibitemShut {NoStop}%
\bibitem [{\citenamefont {Fei}\ \emph {et~al.}(2018)\citenamefont {Fei}, \citenamefont {Huang}, \citenamefont {Malinowski}, \citenamefont {Wang}, \citenamefont {Song}, \citenamefont {Sanchez}, \citenamefont {Yao}, \citenamefont {Xiao}, \citenamefont {Zhu}, \citenamefont {May} \emph {et~al.}}]{fei2018two}%
  \BibitemOpen
  \bibfield  {author} {\bibinfo {author} {\bibfnamefont {Z.}~\bibnamefont {Fei}}, \bibinfo {author} {\bibfnamefont {B.}~\bibnamefont {Huang}}, \bibinfo {author} {\bibfnamefont {P.}~\bibnamefont {Malinowski}}, \bibinfo {author} {\bibfnamefont {W.}~\bibnamefont {Wang}}, \bibinfo {author} {\bibfnamefont {T.}~\bibnamefont {Song}}, \bibinfo {author} {\bibfnamefont {J.}~\bibnamefont {Sanchez}}, \bibinfo {author} {\bibfnamefont {W.}~\bibnamefont {Yao}}, \bibinfo {author} {\bibfnamefont {D.}~\bibnamefont {Xiao}}, \bibinfo {author} {\bibfnamefont {X.}~\bibnamefont {Zhu}}, \bibinfo {author} {\bibfnamefont {A.~F.}\ \bibnamefont {May}}, \emph {et~al.},\ }\bibfield  {title} {\bibinfo {title} {Two-dimensional itinerant ferromagnetism in atomically thin fe3gete2},\ }\href@noop {} {\bibfield  {journal} {\bibinfo  {journal} {Nature materials}\ }\textbf {\bibinfo {volume} {17}},\ \bibinfo {pages} {778} (\bibinfo {year} {2018})}\BibitemShut {NoStop}%
\bibitem [{\citenamefont {Wang}\ \emph {et~al.}(2016)\citenamefont {Wang}, \citenamefont {Du}, \citenamefont {Liu}, \citenamefont {Hu}, \citenamefont {Zhang}, \citenamefont {Zhang}, \citenamefont {Owen}, \citenamefont {Lu}, \citenamefont {Gan}, \citenamefont {Sengupta} \emph {et~al.}}]{wang2016raman}%
  \BibitemOpen
  \bibfield  {author} {\bibinfo {author} {\bibfnamefont {X.}~\bibnamefont {Wang}}, \bibinfo {author} {\bibfnamefont {K.}~\bibnamefont {Du}}, \bibinfo {author} {\bibfnamefont {Y.~Y.~F.}\ \bibnamefont {Liu}}, \bibinfo {author} {\bibfnamefont {P.}~\bibnamefont {Hu}}, \bibinfo {author} {\bibfnamefont {J.}~\bibnamefont {Zhang}}, \bibinfo {author} {\bibfnamefont {Q.}~\bibnamefont {Zhang}}, \bibinfo {author} {\bibfnamefont {M.~H.~S.}\ \bibnamefont {Owen}}, \bibinfo {author} {\bibfnamefont {X.}~\bibnamefont {Lu}}, \bibinfo {author} {\bibfnamefont {C.~K.}\ \bibnamefont {Gan}}, \bibinfo {author} {\bibfnamefont {P.}~\bibnamefont {Sengupta}}, \emph {et~al.},\ }\bibfield  {title} {\bibinfo {title} {Raman spectroscopy of atomically thin two-dimensional magnetic iron phosphorus trisulfide (feps3) crystals},\ }\href@noop {} {\bibfield  {journal} {\bibinfo  {journal} {2D Materials}\ }\textbf {\bibinfo {volume} {3}},\ \bibinfo {pages} {031009} (\bibinfo {year} {2016})}\BibitemShut {NoStop}%
\bibitem [{\citenamefont {Lee}\ \emph {et~al.}(2016)\citenamefont {Lee}, \citenamefont {Lee}, \citenamefont {Ryoo}, \citenamefont {Kang}, \citenamefont {Kim}, \citenamefont {Kim}, \citenamefont {Park}, \citenamefont {Park},\ and\ \citenamefont {Cheong}}]{lee2016ising}%
  \BibitemOpen
  \bibfield  {author} {\bibinfo {author} {\bibfnamefont {J.-U.}\ \bibnamefont {Lee}}, \bibinfo {author} {\bibfnamefont {S.}~\bibnamefont {Lee}}, \bibinfo {author} {\bibfnamefont {J.~H.}\ \bibnamefont {Ryoo}}, \bibinfo {author} {\bibfnamefont {S.}~\bibnamefont {Kang}}, \bibinfo {author} {\bibfnamefont {T.~Y.}\ \bibnamefont {Kim}}, \bibinfo {author} {\bibfnamefont {P.}~\bibnamefont {Kim}}, \bibinfo {author} {\bibfnamefont {C.-H.}\ \bibnamefont {Park}}, \bibinfo {author} {\bibfnamefont {J.-G.}\ \bibnamefont {Park}},\ and\ \bibinfo {author} {\bibfnamefont {H.}~\bibnamefont {Cheong}},\ }\bibfield  {title} {\bibinfo {title} {Ising-type magnetic ordering in atomically thin feps3},\ }\href@noop {} {\bibfield  {journal} {\bibinfo  {journal} {Nano letters}\ }\textbf {\bibinfo {volume} {16}},\ \bibinfo {pages} {7433} (\bibinfo {year} {2016})}\BibitemShut {NoStop}%
\bibitem [{\citenamefont {Song}\ \emph {et~al.}(2022)\citenamefont {Song}, \citenamefont {Occhialini}, \citenamefont {Erge{\c{c}}en}, \citenamefont {Ilyas}, \citenamefont {Amoroso}, \citenamefont {Barone}, \citenamefont {Kapeghian}, \citenamefont {Watanabe}, \citenamefont {Taniguchi}, \citenamefont {Botana}, \citenamefont {Picozzi}, \citenamefont {Gedik},\ and\ \citenamefont {Comin}}]{Song2022}%
  \BibitemOpen
  \bibfield  {author} {\bibinfo {author} {\bibfnamefont {Q.}~\bibnamefont {Song}}, \bibinfo {author} {\bibfnamefont {C.~A.}\ \bibnamefont {Occhialini}}, \bibinfo {author} {\bibfnamefont {E.}~\bibnamefont {Erge{\c{c}}en}}, \bibinfo {author} {\bibfnamefont {B.}~\bibnamefont {Ilyas}}, \bibinfo {author} {\bibfnamefont {D.}~\bibnamefont {Amoroso}}, \bibinfo {author} {\bibfnamefont {P.}~\bibnamefont {Barone}}, \bibinfo {author} {\bibfnamefont {J.}~\bibnamefont {Kapeghian}}, \bibinfo {author} {\bibfnamefont {K.}~\bibnamefont {Watanabe}}, \bibinfo {author} {\bibfnamefont {T.}~\bibnamefont {Taniguchi}}, \bibinfo {author} {\bibfnamefont {A.~S.}\ \bibnamefont {Botana}}, \bibinfo {author} {\bibfnamefont {S.}~\bibnamefont {Picozzi}}, \bibinfo {author} {\bibfnamefont {N.}~\bibnamefont {Gedik}},\ and\ \bibinfo {author} {\bibfnamefont {R.}~\bibnamefont {Comin}},\ }\bibfield  {title} {\bibinfo {title} {{Evidence for a single-layer van der Waals multiferroic}},\ }\href {https://doi.org/10.1038/s41586-021-04337-x} {\bibfield
  {journal} {\bibinfo  {journal} {Nature}\ }\textbf {\bibinfo {volume} {602}},\ \bibinfo {pages} {601} (\bibinfo {year} {2022})}\BibitemShut {NoStop}%
\bibitem [{\citenamefont {Ma}\ \emph {et~al.}(2021)\citenamefont {Ma}, \citenamefont {Hu}, \citenamefont {Li}, \citenamefont {Liu}, \citenamefont {Yao}, \citenamefont {Jia},\ and\ \citenamefont {Liu}}]{ma2021multifunctional}%
  \BibitemOpen
  \bibfield  {author} {\bibinfo {author} {\bibfnamefont {H.-Y.}\ \bibnamefont {Ma}}, \bibinfo {author} {\bibfnamefont {M.}~\bibnamefont {Hu}}, \bibinfo {author} {\bibfnamefont {N.}~\bibnamefont {Li}}, \bibinfo {author} {\bibfnamefont {J.}~\bibnamefont {Liu}}, \bibinfo {author} {\bibfnamefont {W.}~\bibnamefont {Yao}}, \bibinfo {author} {\bibfnamefont {J.-F.}\ \bibnamefont {Jia}},\ and\ \bibinfo {author} {\bibfnamefont {J.}~\bibnamefont {Liu}},\ }\bibfield  {title} {\bibinfo {title} {Multifunctional antiferromagnetic materials with giant piezomagnetism and noncollinear spin current},\ }\href@noop {} {\bibfield  {journal} {\bibinfo  {journal} {Nature communications}\ }\textbf {\bibinfo {volume} {12}},\ \bibinfo {pages} {2846} (\bibinfo {year} {2021})}\BibitemShut {NoStop}%
\bibitem [{\citenamefont {Cui}\ \emph {et~al.}(2023{\natexlab{b}})\citenamefont {Cui}, \citenamefont {Zhu}, \citenamefont {Yao}, \citenamefont {Cui},\ and\ \citenamefont {Yang}}]{cui2023giant}%
  \BibitemOpen
  \bibfield  {author} {\bibinfo {author} {\bibfnamefont {Q.}~\bibnamefont {Cui}}, \bibinfo {author} {\bibfnamefont {Y.}~\bibnamefont {Zhu}}, \bibinfo {author} {\bibfnamefont {X.}~\bibnamefont {Yao}}, \bibinfo {author} {\bibfnamefont {P.}~\bibnamefont {Cui}},\ and\ \bibinfo {author} {\bibfnamefont {H.}~\bibnamefont {Yang}},\ }\bibfield  {title} {\bibinfo {title} {Giant spin-hall and tunneling magnetoresistance effects based on a two-dimensional nonrelativistic antiferromagnetic metal},\ }\href@noop {} {\bibfield  {journal} {\bibinfo  {journal} {Physical Review B}\ }\textbf {\bibinfo {volume} {108}},\ \bibinfo {pages} {024410} (\bibinfo {year} {2023}{\natexlab{b}})}\BibitemShut {NoStop}%
\bibitem [{\citenamefont {Chen}\ \emph {et~al.}(2023)\citenamefont {Chen}, \citenamefont {Wang}, \citenamefont {Li},\ and\ \citenamefont {Sanyal}}]{chen2023giant}%
  \BibitemOpen
  \bibfield  {author} {\bibinfo {author} {\bibfnamefont {X.}~\bibnamefont {Chen}}, \bibinfo {author} {\bibfnamefont {D.}~\bibnamefont {Wang}}, \bibinfo {author} {\bibfnamefont {L.}~\bibnamefont {Li}},\ and\ \bibinfo {author} {\bibfnamefont {B.}~\bibnamefont {Sanyal}},\ }\bibfield  {title} {\bibinfo {title} {Giant spin-splitting and tunable spin-momentum locked transport in room temperature collinear antiferromagnetic semimetallic cro monolayer},\ }\href@noop {} {\bibfield  {journal} {\bibinfo  {journal} {Applied Physics Letters}\ }\textbf {\bibinfo {volume} {123}} (\bibinfo {year} {2023})}\BibitemShut {NoStop}%
\bibitem [{\citenamefont {Guo}\ \emph {et~al.}(2023{\natexlab{a}})\citenamefont {Guo}, \citenamefont {Liu},\ and\ \citenamefont {Lu}}]{guo2023quantum}%
  \BibitemOpen
  \bibfield  {author} {\bibinfo {author} {\bibfnamefont {P.-J.}\ \bibnamefont {Guo}}, \bibinfo {author} {\bibfnamefont {Z.-X.}\ \bibnamefont {Liu}},\ and\ \bibinfo {author} {\bibfnamefont {Z.-Y.}\ \bibnamefont {Lu}},\ }\bibfield  {title} {\bibinfo {title} {Quantum anomalous hall effect in collinear antiferromagnetism},\ }\href@noop {} {\bibfield  {journal} {\bibinfo  {journal} {npj Computational Materials}\ }\textbf {\bibinfo {volume} {9}},\ \bibinfo {pages} {70} (\bibinfo {year} {2023}{\natexlab{a}})}\BibitemShut {NoStop}%
\bibitem [{\citenamefont {Guo}\ \emph {et~al.}(2023{\natexlab{b}})\citenamefont {Guo}, \citenamefont {Guo}, \citenamefont {Cheng}, \citenamefont {Wang},\ and\ \citenamefont {Ang}}]{guo2023piezoelectric}%
  \BibitemOpen
  \bibfield  {author} {\bibinfo {author} {\bibfnamefont {S.-D.}\ \bibnamefont {Guo}}, \bibinfo {author} {\bibfnamefont {X.-S.}\ \bibnamefont {Guo}}, \bibinfo {author} {\bibfnamefont {K.}~\bibnamefont {Cheng}}, \bibinfo {author} {\bibfnamefont {K.}~\bibnamefont {Wang}},\ and\ \bibinfo {author} {\bibfnamefont {Y.~S.}\ \bibnamefont {Ang}},\ }\bibfield  {title} {\bibinfo {title} {Piezoelectric altermagnetism and spin-valley polarization in janus monolayer {Cr}$_2${SO}},\ }\href@noop {} {\bibfield  {journal} {\bibinfo  {journal} {arXiv preprint arXiv:2306.04094}\ } (\bibinfo {year} {2023}{\natexlab{b}})}\BibitemShut {NoStop}%
\bibitem [{\citenamefont {Mermin}\ and\ \citenamefont {Wagner}(1966)}]{mermin1966absence}%
  \BibitemOpen
  \bibfield  {author} {\bibinfo {author} {\bibfnamefont {N.~D.}\ \bibnamefont {Mermin}}\ and\ \bibinfo {author} {\bibfnamefont {H.}~\bibnamefont {Wagner}},\ }\bibfield  {title} {\bibinfo {title} {Absence of ferromagnetism or antiferromagnetism in one-or two-dimensional isotropic heisenberg models},\ }\href@noop {} {\bibfield  {journal} {\bibinfo  {journal} {Physical Review Letters}\ }\textbf {\bibinfo {volume} {17}},\ \bibinfo {pages} {1133} (\bibinfo {year} {1966})}\BibitemShut {NoStop}%
\bibitem [{\citenamefont {Halperin}(2019)}]{halperin2019hohenberg}%
  \BibitemOpen
  \bibfield  {author} {\bibinfo {author} {\bibfnamefont {B.~I.}\ \bibnamefont {Halperin}},\ }\bibfield  {title} {\bibinfo {title} {On the hohenberg--mermin--wagner theorem and its limitations},\ }\href@noop {} {\bibfield  {journal} {\bibinfo  {journal} {Journal of Statistical Physics}\ }\textbf {\bibinfo {volume} {175}},\ \bibinfo {pages} {521} (\bibinfo {year} {2019})}\BibitemShut {NoStop}%
\bibitem [{\citenamefont {{\v{S}}mejkal}\ \emph {et~al.}(2022{\natexlab{c}})\citenamefont {{\v{S}}mejkal}, \citenamefont {Sinova},\ and\ \citenamefont {Jungwirth}}]{vsmejkal2022beyond}%
  \BibitemOpen
  \bibfield  {author} {\bibinfo {author} {\bibfnamefont {L.}~\bibnamefont {{\v{S}}mejkal}}, \bibinfo {author} {\bibfnamefont {J.}~\bibnamefont {Sinova}},\ and\ \bibinfo {author} {\bibfnamefont {T.}~\bibnamefont {Jungwirth}},\ }\bibfield  {title} {\bibinfo {title} {Beyond conventional ferromagnetism and antiferromagnetism: A phase with nonrelativistic spin and crystal rotation symmetry},\ }\href@noop {} {\bibfield  {journal} {\bibinfo  {journal} {Physical Review X}\ }\textbf {\bibinfo {volume} {12}},\ \bibinfo {pages} {031042} (\bibinfo {year} {2022}{\natexlab{c}})}\BibitemShut {NoStop}%
\bibitem [{\citenamefont {Gjerding}\ \emph {et~al.}(2021)\citenamefont {Gjerding}, \citenamefont {Taghizadeh}, \citenamefont {Rasmussen}, \citenamefont {Ali}, \citenamefont {Bertoldo}, \citenamefont {Deilmann}, \citenamefont {Kn{\o}sgaard}, \citenamefont {Kruse}, \citenamefont {Larsen}, \citenamefont {Manti}, \citenamefont {Pedersen}, \citenamefont {Petralanda}, \citenamefont {Skovhus}, \citenamefont {Svendsen}, \citenamefont {Mortensen}, \citenamefont {Olsen},\ and\ \citenamefont {Thygesen}}]{Gjerding2021}%
  \BibitemOpen
  \bibfield  {author} {\bibinfo {author} {\bibfnamefont {M.~N.}\ \bibnamefont {Gjerding}}, \bibinfo {author} {\bibfnamefont {A.}~\bibnamefont {Taghizadeh}}, \bibinfo {author} {\bibfnamefont {A.}~\bibnamefont {Rasmussen}}, \bibinfo {author} {\bibfnamefont {S.}~\bibnamefont {Ali}}, \bibinfo {author} {\bibfnamefont {F.}~\bibnamefont {Bertoldo}}, \bibinfo {author} {\bibfnamefont {T.}~\bibnamefont {Deilmann}}, \bibinfo {author} {\bibfnamefont {N.~R.}\ \bibnamefont {Kn{\o}sgaard}}, \bibinfo {author} {\bibfnamefont {M.}~\bibnamefont {Kruse}}, \bibinfo {author} {\bibfnamefont {A.~H.}\ \bibnamefont {Larsen}}, \bibinfo {author} {\bibfnamefont {S.}~\bibnamefont {Manti}}, \bibinfo {author} {\bibfnamefont {T.~G.}\ \bibnamefont {Pedersen}}, \bibinfo {author} {\bibfnamefont {U.}~\bibnamefont {Petralanda}}, \bibinfo {author} {\bibfnamefont {T.}~\bibnamefont {Skovhus}}, \bibinfo {author} {\bibfnamefont {M.~K.}\ \bibnamefont {Svendsen}}, \bibinfo {author} {\bibfnamefont {J.~J.}\ \bibnamefont {Mortensen}}, \bibinfo {author}
  {\bibfnamefont {T.}~\bibnamefont {Olsen}},\ and\ \bibinfo {author} {\bibfnamefont {K.~S.}\ \bibnamefont {Thygesen}},\ }\bibfield  {title} {\bibinfo {title} {{Recent progress of the computational 2D materials database (C2DB)}},\ }\href {https://doi.org/10.1088/2053-1583/ac1059} {\bibfield  {journal} {\bibinfo  {journal} {2D Mater.}\ }\textbf {\bibinfo {volume} {8}},\ \bibinfo {pages} {044002} (\bibinfo {year} {2021})},\ \Eprint {https://arxiv.org/abs/2102.03029} {2102.03029} \BibitemShut {NoStop}%
\bibitem [{\citenamefont {S{\o}dequist}\ and\ \citenamefont {Olsen}(2023{\natexlab{a}})}]{sodequist2023magnetic}%
  \BibitemOpen
  \bibfield  {author} {\bibinfo {author} {\bibfnamefont {J.}~\bibnamefont {S{\o}dequist}}\ and\ \bibinfo {author} {\bibfnamefont {T.}~\bibnamefont {Olsen}},\ }\bibfield  {title} {\bibinfo {title} {Magnetic order in the computational 2d materials database (c2db) from high throughput spin spiral calculations},\ }\href@noop {} {\bibfield  {journal} {\bibinfo  {journal} {arXiv preprint arXiv:2309.11945}\ } (\bibinfo {year} {2023}{\natexlab{a}})}\BibitemShut {NoStop}%
\bibitem [{\citenamefont {Torelli}\ and\ \citenamefont {Olsen}(2020)}]{torelli2020first}%
  \BibitemOpen
  \bibfield  {author} {\bibinfo {author} {\bibfnamefont {D.}~\bibnamefont {Torelli}}\ and\ \bibinfo {author} {\bibfnamefont {T.}~\bibnamefont {Olsen}},\ }\bibfield  {title} {\bibinfo {title} {First principles heisenberg models of 2d magnetic materials: the importance of quantum corrections to the exchange coupling},\ }\href@noop {} {\bibfield  {journal} {\bibinfo  {journal} {Journal of Physics: Condensed Matter}\ }\textbf {\bibinfo {volume} {32}},\ \bibinfo {pages} {335802} (\bibinfo {year} {2020})}\BibitemShut {NoStop}%
\bibitem [{\citenamefont {Togo}\ and\ \citenamefont {Tanaka}(2018)}]{spglibv1}%
  \BibitemOpen
  \bibfield  {author} {\bibinfo {author} {\bibfnamefont {A.}~\bibnamefont {Togo}}\ and\ \bibinfo {author} {\bibfnamefont {I.}~\bibnamefont {Tanaka}},\ }\href@noop {} {\bibinfo {title} {$\texttt{Spglib}$: a software library for crystal symmetry search}},\ \bibinfo {howpublished} {\url{https://github.com/spglib/spglib}} (\bibinfo {year} {2018}),\ \Eprint {https://arxiv.org/abs/arXiv:1808.01590} {arXiv:1808.01590} \BibitemShut {NoStop}%
\bibitem [{\citenamefont {Shinohara}\ \emph {et~al.}(2023)\citenamefont {Shinohara}, \citenamefont {Togo},\ and\ \citenamefont {Tanaka}}]{spglibv2}%
  \BibitemOpen
  \bibfield  {author} {\bibinfo {author} {\bibfnamefont {K.}~\bibnamefont {Shinohara}}, \bibinfo {author} {\bibfnamefont {A.}~\bibnamefont {Togo}},\ and\ \bibinfo {author} {\bibfnamefont {I.}~\bibnamefont {Tanaka}},\ }\bibfield  {title} {\bibinfo {title} {{Algorithms for magnetic symmetry operation search and identification of magnetic space group from magnetic crystal structure}},\ }\href {https://doi.org/10.1107/S2053273323005016} {\bibfield  {journal} {\bibinfo  {journal} {Acta Cryst. A}\ }\textbf {\bibinfo {volume} {79}},\ \bibinfo {pages} {390} (\bibinfo {year} {2023})}\BibitemShut {NoStop}%
\bibitem [{\citenamefont {S{\o}dequist}\ and\ \citenamefont {Olsen}(2023{\natexlab{b}})}]{sodequist2023type}%
  \BibitemOpen
  \bibfield  {author} {\bibinfo {author} {\bibfnamefont {J.}~\bibnamefont {S{\o}dequist}}\ and\ \bibinfo {author} {\bibfnamefont {T.}~\bibnamefont {Olsen}},\ }\bibfield  {title} {\bibinfo {title} {Type ii multiferroic order in two-dimensional transition metal halides from first principles spin-spiral calculations},\ }\href@noop {} {\bibfield  {journal} {\bibinfo  {journal} {2D Materials}\ }\textbf {\bibinfo {volume} {10}},\ \bibinfo {pages} {035016} (\bibinfo {year} {2023}{\natexlab{b}})}\BibitemShut {NoStop}%
\bibitem [{\citenamefont {Kurz}\ \emph {et~al.}(2004)\citenamefont {Kurz}, \citenamefont {F{\"o}rster}, \citenamefont {Nordstr{\"o}m}, \citenamefont {Bihlmayer},\ and\ \citenamefont {Bl{\"u}gel}}]{kurz2004ab}%
  \BibitemOpen
  \bibfield  {author} {\bibinfo {author} {\bibfnamefont {P.}~\bibnamefont {Kurz}}, \bibinfo {author} {\bibfnamefont {F.}~\bibnamefont {F{\"o}rster}}, \bibinfo {author} {\bibfnamefont {L.}~\bibnamefont {Nordstr{\"o}m}}, \bibinfo {author} {\bibfnamefont {G.}~\bibnamefont {Bihlmayer}},\ and\ \bibinfo {author} {\bibfnamefont {S.}~\bibnamefont {Bl{\"u}gel}},\ }\bibfield  {title} {\bibinfo {title} {Ab initio treatment of noncollinear magnets with the full-potential linearized augmented plane wave method},\ }\href@noop {} {\bibfield  {journal} {\bibinfo  {journal} {Physical Review B}\ }\textbf {\bibinfo {volume} {69}},\ \bibinfo {pages} {024415} (\bibinfo {year} {2004})}\BibitemShut {NoStop}%
\bibitem [{\citenamefont {Kn{\"o}pfle}\ \emph {et~al.}(2000)\citenamefont {Kn{\"o}pfle}, \citenamefont {Sandratskii},\ and\ \citenamefont {K{\"u}bler}}]{knopfle2000spin}%
  \BibitemOpen
  \bibfield  {author} {\bibinfo {author} {\bibfnamefont {K.}~\bibnamefont {Kn{\"o}pfle}}, \bibinfo {author} {\bibfnamefont {L.}~\bibnamefont {Sandratskii}},\ and\ \bibinfo {author} {\bibfnamefont {J.}~\bibnamefont {K{\"u}bler}},\ }\bibfield  {title} {\bibinfo {title} {Spin spiral ground state of $\gamma$-iron},\ }\href@noop {} {\bibfield  {journal} {\bibinfo  {journal} {Physical Review B}\ }\textbf {\bibinfo {volume} {62}},\ \bibinfo {pages} {5564} (\bibinfo {year} {2000})}\BibitemShut {NoStop}%
\bibitem [{\citenamefont {Enkovaara}\ \emph {et~al.}(2010)\citenamefont {Enkovaara}, \citenamefont {Rostgaard}, \citenamefont {Mortensen}, \citenamefont {Chen}, \citenamefont {Du{\l}ak}, \citenamefont {Ferrighi}, \citenamefont {Gavnholt}, \citenamefont {Glinsvad}, \citenamefont {Haikola}, \citenamefont {Hansen}, \citenamefont {Kristoffersen}, \citenamefont {Kuisma}, \citenamefont {Larsen}, \citenamefont {Lehtovaara}, \citenamefont {Ljungberg}, \citenamefont {Lopez-Acevedo}, \citenamefont {Moses}, \citenamefont {Ojanen}, \citenamefont {Olsen}, \citenamefont {Petzold}, \citenamefont {Romero}, \citenamefont {Stausholm-M{\o}ller}, \citenamefont {Strange}, \citenamefont {Tritsaris}, \citenamefont {Vanin}, \citenamefont {Walter}, \citenamefont {Hammer}, \citenamefont {H{\"{a}}kkinen}, \citenamefont {Madsen}, \citenamefont {Nieminen}, \citenamefont {N{\o}rskov}, \citenamefont {Puska}, \citenamefont {Rantala}, \citenamefont {Schi{\o}tz}, \citenamefont {Thygesen}, \citenamefont {Jacobsen},\ and\ \citenamefont
  {Others}}]{Enkovaara2010}%
  \BibitemOpen
  \bibfield  {author} {\bibinfo {author} {\bibfnamefont {J.}~\bibnamefont {Enkovaara}}, \bibinfo {author} {\bibfnamefont {C.}~\bibnamefont {Rostgaard}}, \bibinfo {author} {\bibfnamefont {J.~J.}\ \bibnamefont {Mortensen}}, \bibinfo {author} {\bibfnamefont {J.}~\bibnamefont {Chen}}, \bibinfo {author} {\bibfnamefont {M.}~\bibnamefont {Du{\l}ak}}, \bibinfo {author} {\bibfnamefont {L.}~\bibnamefont {Ferrighi}}, \bibinfo {author} {\bibfnamefont {J.}~\bibnamefont {Gavnholt}}, \bibinfo {author} {\bibfnamefont {C.}~\bibnamefont {Glinsvad}}, \bibinfo {author} {\bibfnamefont {V.}~\bibnamefont {Haikola}}, \bibinfo {author} {\bibfnamefont {H.~a.}\ \bibnamefont {Hansen}}, \bibinfo {author} {\bibfnamefont {H.~H.}\ \bibnamefont {Kristoffersen}}, \bibinfo {author} {\bibfnamefont {M.}~\bibnamefont {Kuisma}}, \bibinfo {author} {\bibfnamefont {a.~H.}\ \bibnamefont {Larsen}}, \bibinfo {author} {\bibfnamefont {L.}~\bibnamefont {Lehtovaara}}, \bibinfo {author} {\bibfnamefont {M.}~\bibnamefont {Ljungberg}}, \bibinfo {author}
  {\bibfnamefont {O.}~\bibnamefont {Lopez-Acevedo}}, \bibinfo {author} {\bibfnamefont {P.~G.}\ \bibnamefont {Moses}}, \bibinfo {author} {\bibfnamefont {J.}~\bibnamefont {Ojanen}}, \bibinfo {author} {\bibfnamefont {T.}~\bibnamefont {Olsen}}, \bibinfo {author} {\bibfnamefont {V.}~\bibnamefont {Petzold}}, \bibinfo {author} {\bibfnamefont {N.~a.}\ \bibnamefont {Romero}}, \bibinfo {author} {\bibfnamefont {J.}~\bibnamefont {Stausholm-M{\o}ller}}, \bibinfo {author} {\bibfnamefont {M.}~\bibnamefont {Strange}}, \bibinfo {author} {\bibfnamefont {G.~a.}\ \bibnamefont {Tritsaris}}, \bibinfo {author} {\bibfnamefont {M.}~\bibnamefont {Vanin}}, \bibinfo {author} {\bibfnamefont {M.}~\bibnamefont {Walter}}, \bibinfo {author} {\bibfnamefont {B.}~\bibnamefont {Hammer}}, \bibinfo {author} {\bibfnamefont {H.}~\bibnamefont {H{\"{a}}kkinen}}, \bibinfo {author} {\bibfnamefont {G.~K.~H.}\ \bibnamefont {Madsen}}, \bibinfo {author} {\bibfnamefont {R.~M.}\ \bibnamefont {Nieminen}}, \bibinfo {author} {\bibfnamefont {J.~K.}\ \bibnamefont
  {N{\o}rskov}}, \bibinfo {author} {\bibfnamefont {M.}~\bibnamefont {Puska}}, \bibinfo {author} {\bibfnamefont {T.~T.}\ \bibnamefont {Rantala}}, \bibinfo {author} {\bibfnamefont {J.}~\bibnamefont {Schi{\o}tz}}, \bibinfo {author} {\bibfnamefont {K.~S.}\ \bibnamefont {Thygesen}}, \bibinfo {author} {\bibfnamefont {K.~W.}\ \bibnamefont {Jacobsen}},\ and\ \bibinfo {author} {\bibnamefont {Others}},\ }\bibfield  {title} {\bibinfo {title} {{Electronic structure calculations with GPAW: a real-space implementation of the projector augmented-wave method}},\ }\href {https://doi.org/10.1088/0953-8984/22/25/253202} {\bibfield  {journal} {\bibinfo  {journal} {J. Phys. Condens. Matter}\ }\textbf {\bibinfo {volume} {22}},\ \bibinfo {pages} {253202} (\bibinfo {year} {2010})}\BibitemShut {NoStop}%
\bibitem [{\citenamefont {Mortensen}\ \emph {et~al.}(2023)\citenamefont {Mortensen}, \citenamefont {Larsen}, \citenamefont {Kuisma}, \citenamefont {Ivanov}, \citenamefont {Taghizadeh}, \citenamefont {Peterson}, \citenamefont {Haldar}, \citenamefont {Dohn}, \citenamefont {Sch{\"a}fer}, \citenamefont {J{\'o}nsson} \emph {et~al.}}]{mortensen2023gpaw}%
  \BibitemOpen
  \bibfield  {author} {\bibinfo {author} {\bibfnamefont {J.~J.}\ \bibnamefont {Mortensen}}, \bibinfo {author} {\bibfnamefont {A.~H.}\ \bibnamefont {Larsen}}, \bibinfo {author} {\bibfnamefont {M.}~\bibnamefont {Kuisma}}, \bibinfo {author} {\bibfnamefont {A.~V.}\ \bibnamefont {Ivanov}}, \bibinfo {author} {\bibfnamefont {A.}~\bibnamefont {Taghizadeh}}, \bibinfo {author} {\bibfnamefont {A.}~\bibnamefont {Peterson}}, \bibinfo {author} {\bibfnamefont {A.}~\bibnamefont {Haldar}}, \bibinfo {author} {\bibfnamefont {A.~O.}\ \bibnamefont {Dohn}}, \bibinfo {author} {\bibfnamefont {C.}~\bibnamefont {Sch{\"a}fer}}, \bibinfo {author} {\bibfnamefont {E.~{\"O}.}\ \bibnamefont {J{\'o}nsson}}, \emph {et~al.},\ }\bibfield  {title} {\bibinfo {title} {Gpaw: open python package for electronic-structure calculations},\ }\href@noop {} {\bibfield  {journal} {\bibinfo  {journal} {arXiv preprint arXiv:2310.14776}\ } (\bibinfo {year} {2023})}\BibitemShut {NoStop}%
\bibitem [{\citenamefont {Allmann}\ and\ \citenamefont {Hinek}(2007)}]{Allmann2007}%
  \BibitemOpen
  \bibfield  {author} {\bibinfo {author} {\bibfnamefont {R.}~\bibnamefont {Allmann}}\ and\ \bibinfo {author} {\bibfnamefont {R.}~\bibnamefont {Hinek}},\ }\bibfield  {title} {\bibinfo {title} {{The introduction of structure types into the Inorganic Crystal Structure Database ICSD}},\ }\href {https://doi.org/10.1107/S0108767307038081} {\bibfield  {journal} {\bibinfo  {journal} {Acta Crystallogr. Sect. A Found. Crystallogr.}\ }\textbf {\bibinfo {volume} {63}},\ \bibinfo {pages} {412} (\bibinfo {year} {2007})}\BibitemShut {NoStop}%
\bibitem [{\citenamefont {Gra{\v{z}}ulis}\ \emph {et~al.}(2011)\citenamefont {Gra{\v{z}}ulis}, \citenamefont {Da{\v{s}}kevi{\v{c}}}, \citenamefont {Merkys}, \citenamefont {Chateigner}, \citenamefont {Lutterotti}, \citenamefont {Quir{\'{o}}s}, \citenamefont {Serebryanaya}, \citenamefont {Moeck}, \citenamefont {Downs},\ and\ \citenamefont {Bail}}]{Graulis2011}%
  \BibitemOpen
  \bibfield  {author} {\bibinfo {author} {\bibfnamefont {S.}~\bibnamefont {Gra{\v{z}}ulis}}, \bibinfo {author} {\bibfnamefont {A.}~\bibnamefont {Da{\v{s}}kevi{\v{c}}}}, \bibinfo {author} {\bibfnamefont {A.}~\bibnamefont {Merkys}}, \bibinfo {author} {\bibfnamefont {D.}~\bibnamefont {Chateigner}}, \bibinfo {author} {\bibfnamefont {L.}~\bibnamefont {Lutterotti}}, \bibinfo {author} {\bibfnamefont {M.}~\bibnamefont {Quir{\'{o}}s}}, \bibinfo {author} {\bibfnamefont {N.~R.}\ \bibnamefont {Serebryanaya}}, \bibinfo {author} {\bibfnamefont {P.}~\bibnamefont {Moeck}}, \bibinfo {author} {\bibfnamefont {R.~T.}\ \bibnamefont {Downs}},\ and\ \bibinfo {author} {\bibfnamefont {A.~L.}\ \bibnamefont {Bail}},\ }\bibfield  {title} {\bibinfo {title} {Crystallography open database ({COD}): an open-access collection of crystal structures and platform for world-wide collaboration},\ }\href {https://doi.org/10.1093/nar/gkr900} {\bibfield  {journal} {\bibinfo  {journal} {Nucleic Acids Research}\ }\textbf {\bibinfo {volume} {40}},\
  \bibinfo {pages} {D420} (\bibinfo {year} {2011})}\BibitemShut {NoStop}%
\bibitem [{\citenamefont {Young}\ and\ \citenamefont {Kane}(2015)}]{young2015dirac}%
  \BibitemOpen
  \bibfield  {author} {\bibinfo {author} {\bibfnamefont {S.~M.}\ \bibnamefont {Young}}\ and\ \bibinfo {author} {\bibfnamefont {C.~L.}\ \bibnamefont {Kane}},\ }\bibfield  {title} {\bibinfo {title} {Dirac semimetals in two dimensions},\ }\href@noop {} {\bibfield  {journal} {\bibinfo  {journal} {Physical review letters}\ }\textbf {\bibinfo {volume} {115}},\ \bibinfo {pages} {126803} (\bibinfo {year} {2015})}\BibitemShut {NoStop}%
\bibitem [{\citenamefont {Zhao}\ and\ \citenamefont {Schnyder}(2016)}]{zhao2016nonsymmorphic}%
  \BibitemOpen
  \bibfield  {author} {\bibinfo {author} {\bibfnamefont {Y.}~\bibnamefont {Zhao}}\ and\ \bibinfo {author} {\bibfnamefont {A.~P.}\ \bibnamefont {Schnyder}},\ }\bibfield  {title} {\bibinfo {title} {Nonsymmorphic symmetry-required band crossings in topological semimetals},\ }\href@noop {} {\bibfield  {journal} {\bibinfo  {journal} {Physical Review B}\ }\textbf {\bibinfo {volume} {94}},\ \bibinfo {pages} {195109} (\bibinfo {year} {2016})}\BibitemShut {NoStop}%
\bibitem [{\citenamefont {Durhuus}\ \emph {et~al.}(2023)\citenamefont {Durhuus}, \citenamefont {Skovhus},\ and\ \citenamefont {Olsen}}]{Durhuus2023}%
  \BibitemOpen
  \bibfield  {author} {\bibinfo {author} {\bibfnamefont {F.~L.}\ \bibnamefont {Durhuus}}, \bibinfo {author} {\bibfnamefont {T.}~\bibnamefont {Skovhus}},\ and\ \bibinfo {author} {\bibfnamefont {T.}~\bibnamefont {Olsen}},\ }\bibfield  {title} {\bibinfo {title} {{Plane wave implementation of the magnetic force theorem for magnetic exchange constants: application to bulk Fe, Co and Ni}},\ }\href {https://doi.org/10.1088/1361-648X/acab4b} {\bibfield  {journal} {\bibinfo  {journal} {Journal of Physics: Condensed Matter}\ }\textbf {\bibinfo {volume} {35}},\ \bibinfo {pages} {105802} (\bibinfo {year} {2023})},\ \Eprint {https://arxiv.org/abs/2204.04169} {2204.04169} \BibitemShut {NoStop}%
\bibitem [{\citenamefont {Skovhus}\ and\ \citenamefont {Olsen}(2022)}]{Skovhus2022}%
  \BibitemOpen
  \bibfield  {author} {\bibinfo {author} {\bibfnamefont {T.}~\bibnamefont {Skovhus}}\ and\ \bibinfo {author} {\bibfnamefont {T.}~\bibnamefont {Olsen}},\ }\bibfield  {title} {\bibinfo {title} {{Magnons in antiferromagnetic bcc Cr and Cr$_2$O$_3$ from time-dependent density functional theory}},\ }\href {https://doi.org/10.1103/PhysRevB.106.085131} {\bibfield  {journal} {\bibinfo  {journal} {Physical Review B}\ }\textbf {\bibinfo {volume} {106}},\ \bibinfo {pages} {085131} (\bibinfo {year} {2022})}\BibitemShut {NoStop}%
\bibitem [{\citenamefont {Liechtenstein}\ \emph {et~al.}(1987)\citenamefont {Liechtenstein}, \citenamefont {Katsnelson}, \citenamefont {Antropov},\ and\ \citenamefont {Gubanov}}]{Liechtenstein1987}%
  \BibitemOpen
  \bibfield  {author} {\bibinfo {author} {\bibfnamefont {A.}~\bibnamefont {Liechtenstein}}, \bibinfo {author} {\bibfnamefont {M.}~\bibnamefont {Katsnelson}}, \bibinfo {author} {\bibfnamefont {V.}~\bibnamefont {Antropov}},\ and\ \bibinfo {author} {\bibfnamefont {V.}~\bibnamefont {Gubanov}},\ }\bibfield  {title} {\bibinfo {title} {{Local spin density functional approach to the theory of exchange interactions in ferromagnetic metals and alloys}},\ }\href {https://doi.org/10.1016/0304-8853(87)90721-9} {\bibfield  {journal} {\bibinfo  {journal} {Journal of Magnetism and Magnetic Materials}\ }\textbf {\bibinfo {volume} {67}},\ \bibinfo {pages} {65} (\bibinfo {year} {1987})}\BibitemShut {NoStop}%
\bibitem [{\citenamefont {Olsen}(2016)}]{Olsen2016a}%
  \BibitemOpen
  \bibfield  {author} {\bibinfo {author} {\bibfnamefont {T.}~\bibnamefont {Olsen}},\ }\bibfield  {title} {\bibinfo {title} {{Designing in-plane heterostructures of quantum spin Hall insulators from first principles: 1T'-MoS$_2$ with adsorbates}},\ }\href {https://doi.org/10.1103/PhysRevB.94.235106} {\bibfield  {journal} {\bibinfo  {journal} {Physical Review B}\ }\textbf {\bibinfo {volume} {94}},\ \bibinfo {pages} {235106} (\bibinfo {year} {2016})},\ \Eprint {https://arxiv.org/abs/1609.02338} {1609.02338} \BibitemShut {NoStop}%
\end{thebibliography}%

\end{document}